\documentclass[runningheads]{llncs}
\usepackage[T1]{fontenc}
\usepackage{graphicx}
\usepackage{float}
\usepackage{subfiles}
\usepackage{xcolor}
\usepackage[colorlinks=true,linkcolor=blue,citecolor=blue]{hyperref}
\usepackage{cleveref}
\usepackage{cprotect}
\usepackage{cite}
\usepackage{minted}
\usepackage{listings}
\usepackage{orcidlink}

\newif\ifdraft\drafttrue

\newif\iffull\fulltrue
\begin{document}

\newcommand{\Tool}{\textsc{DrvHorn}}
\newcommand\TODO[1]{\textcolor{orange}{TODO: #1}}

\setlength{\emergencystretch}{3em}

\graphicspath{ {./images/} }

\lstset{
  basicstyle=\ttfamily,
  columns=fullflexible,
  breaklines=true,
  breakatwhitespace,
}
\newcommand\code[1]{\lstinline|#1|}

\definecolor{highlightred}{RGB}{250,180,180}
\newcommand{\diffadd}[1]{\textcolor{green}{#1}}
\newcommand{\diffrem}[1]{\textcolor{red}{#1}}

\newcommand\totalPlatformDrivers{3387}
\newcommand\timeouts{807}
\newcommand\timeoutsPerc{23.8\%}
\newcommand\allBugs{777}
\newcommand\realBugs{545}
\newcommand\unknownBugs{424}
\newcommand\falsePositiveRate{29.9\%}
\newcommand\falsePositiveCount{232}
\newcommand\falsePositiveByMemoryAnalysis{205}
\newcommand\falsePositiveBySlicing{27}
\newcommand\timeoutByMtkClk{156}

\newcommand\mergedPatches{45}
\newcommand\memLeakPatches{41}

\newcommand\ofThermalZoneFind{113}
\newcommand\oppAddStatic{88}
\newcommand\devmWakeupUsers{31}

\newcommand\bmcErrorCountInDetected{215}
\newcommand\bmcErrorRateInDetected{39.4\%}
\newcommand\bmcTimeoutCountInDetected{132}
\newcommand\bmcTimeoutRateInDetected{24.2\%}

\newcommand\bmcTimeoutInTimeout{372}
\newcommand\bmcBugsInTimeout{37}
\newcommand\bmcCorrectBugsInTimeout{14}
\newcommand\bmcFPBugsInTimeout{23}

\crefformat{section}{Section~#2#1#3}
\crefformat{subsection}{Section~#2#1#3}
\crefformat{subsubsection}{Section~#2#1#3}

\title{Automatic Detection of Reference Counting Bugs in Linux Kernel Drivers}
\titlerunning{Automatic Detection of Reference Counting Bugs in Linux Kernel Drivers}

\author{
  Joe Hattori\thanks{now at Google}\orcidlink{0000-0001-5177-8275}
  \and Naoki Kobayashi\orcidlink{0000-0002-0537-0604}
  \and Ken Sakayori\orcidlink{0000-0003-3238-9279}
}

\institute{The University of Tokyo
\email{\{joe,koba,sakayori\}@is.s.u-tokyo.ac.jp}}

\maketitle

\usemintedstyle{colorful}

\begin{abstract}
  Reference counting bugs in Linux kernel drivers can lead to severe resource mismanagement and security vulnerabilities.
We introduce \Tool, a novel automated tool to detect these bugs by reducing reference counting verification to an assertion checking problem leveraging the Linux driver interface.
Through efficient modeling of the Linux kernel and aggressive program slicing, \Tool\ discovered \realBugs\ bugs, of which \unknownBugs\ were previously unknown, across all platform drivers in v6.6 Linux kernel, with a lower false positive rate of \falsePositiveRate\ compared to prior studies.
To address the root causes of these newly discovered bugs, we submitted patches to the Linux kernel, and \mergedPatches\ of them were merged.

\end{abstract}

\section{Introduction} \label{sec:introduction}
In a monolithic OS kernel like Linux, device drivers operate at the same privilege level as the rest of the kernel, making any bugs in drivers a substantial security risk.
This concern is amplified by the size of the driver codebase---approximately 66.5\% of the entire kernel,\footnote{Based on lines of code in the initial v6.6 release}---and by the fact that users can develop and load their own drivers at will.
Given the above, static and automated verification of Linux drivers' bug-freeness is crucial.

The Linux kernel is predominantly written in C and reference counts (refcounts) are extensively used to manually manage the lifecycle of kernel objects~\cite{kroah2004kobjects}.
A refcount tracks the number of references to an object, where a positive refcount means the object is in use, and zero means it is no longer used and can be freed.
Refcount mechanisms are used extensively across all parts of the kernel, including drivers.

Refcount bugs occur when the refcount does not match the actual number of references~\cite{emmi2009verifying}.
Excessive refcount causes memory leaks, and insufficient refcount leads to use-after-free (UAF).
CVE-2023-7192~\cite{cve-2023-7192} is an example of a refcount bug that leaks memory, allowing a local attacker with \code{CAP_NET_ADMIN} to perform a DoS attack.
Many other refcount bugs with varying impact have also been reported in the Linux kernel~\cite{cve-2019-11487,bai2024countdown}.

We present \Tool, a novel tool to automatically detect refcount bugs in Linux drivers\footnote{\Tool~is available at \url{https://github.com/joehattori/drvhorn}}.
By focusing on drivers, \Tool\ achieves scalable refcount bug detection with high accuracy.
\Tool\ utilizes the kernel-provided driver interface to reduce refcount verification into an assertion checking problem.
\Tool\ can be integrated with arbitrary tools that perform assertion checking on LLVM bitcode.
In our evaluation, we employed SeaHorn~\cite{gurfinkel2015seahorn}, a state-of-the-art CHC-based verification framework, to analyze all platform drivers in x86-64 Linux v6.6.
It reported \allBugs\ bugs, of which \realBugs\ bugs were real, showing a relatively low false positive rate of \falsePositiveRate\ compared to prior studies.
Out of the real bugs, \unknownBugs\ were previously unknown, and \mergedPatches\ patches by us were merged into the Linux kernel to address the root causes of these newly discovered bugs.

The structure of this paper is as follows.
\cref{sec:tool-overview} provides an overview of \Tool.
\cref{sec:implementation} explains the key techniques of our approach.
\cref{sec:evaluation} presents the evaluation results.
\cref{sec:related-work} discusses the related work, and \cref{sec:conclusion} concludes the paper with future work.

\section{Overview} \label{sec:tool-overview}
This section provides a brief overview of \Tool{} and illustrates its capabilities with a concrete example.
\Tool{} takes two inputs: the Linux kernel source code compiled to LLVM bitcode and the name of the target driver.
The tool then analyzes the inputs fully automatically and outputs whether or not there may exist a refcount bug in the initialization phase of the target driver.\footnote{Producing a more informative report is left for future work.}
\Tool\ requires only three kernel modifications: Makefile support for generating the LLVM bitcode, adding a dummy field to \code{struct kref}\footnote{This is done to prevent LLVM to merge structurally equivalent types}, and disabling inlining of refcount APIs such as \code{kref_get} and \code{kref_put}.

Refcount manipulations primarily occur in initialization and cleanup functions.
Since the main responsibility of cleanup functions is releasing resources while error paths of initialization functions are prone to missing it, any invalid refcount manipulation in the cleanup path would most likely also exist in the error path of the initialization function.
Therefore, \Tool\ focuses on the initialization phase, and our evaluation confirms the effectiveness of this approach by discovering \realBugs\ bugs including bugs in cleanup functions\footnote{Commit 74adad5 found a refcount bug which also resided in a cleanup function}.
The analysis of other parts is left for future work.
Extending \Tool\ to termination/release functions should require only modest implementation effort because the same modeling, slicing, and assertion-checking pipeline can be reused.
The main challenge is that analyzing more driver code and more environment interaction can increase both false positives and false negatives.
The practical benefit may also be limited, since many release-function refcount bugs would likely already be reflected in initialization error paths.
\Tool\ is neither sound nor complete: it may miss a bug and also report a false bug.
However, our evaluation suggests that the false negative/positive rates are small.

Figure~\ref{fig:example-refcount-bug} shows an example of a real-world refcount bug found and fixed by using \Tool\ in commit e386209.
\code{bcm_sf2_mdio_register()} iterates over child nodes using \code{for_each_available_child_of_node()}.
Within the loop, \code{of_phy_find_device()} increments the refcount of \code{struct phy_device}, which is then removed by \code{phy_device_remove()}.
However, the incremented refcount was not decremented, causing a memory leak that was fixed by adding \code{phy_device_free()}.

\begin{figure}[h!]
\begin{minted}[escapeinside=@@]{c}
#define for_each_available_child_of_node(parent, child) \
  for (child = of_get_next_available_child(parent, NULL); \
    child != NULL; child = of_get_next_available_child(parent, child))

static int bcm_sf2_mdio_register(struct dsa_switch *ds) {
  struct device_node *dn, *child;
  ...
  for_each_available_child_of_node(dn, child) {
    ...
    struct phy_device *phydev = of_phy_find_device(child);
    if (phydev) {
      phy_device_remove(phydev);
@\diffadd{+     phy\_device\_free(phydev);}@
    }
  }
  ...
}
\end{minted}
\caption{Commit e386209}
\label{fig:example-refcount-bug}
\end{figure}

\section{Proposed Method} \label{sec:implementation}
\begin{figure}[ht]
\centering
\includegraphics[
  width=\linewidth,
  alt={Architecture diagram of DrvHorn. The Linux kernel bitcode is modeled, and along with the target driver name, refcount bug detection is reduced to an assertion problem. After slicing, the assertion problem is fed to an off-the-shelf solver, and DrvHorn obtains the refcount verification result.}
]{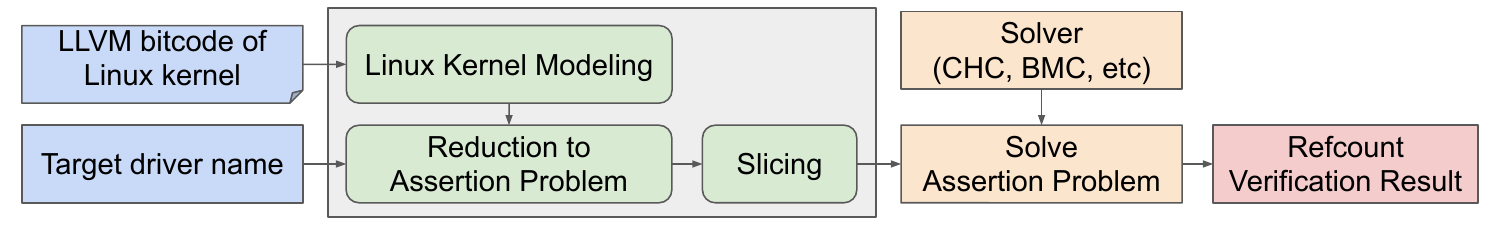}
\caption{Overall architecture of \Tool}
\label{fig:tool-overview}
\end{figure}

Figure~\ref{fig:tool-overview} shows the overall architecture of \Tool.
\Tool\ transforms the refcount verification problem into an assertion problem (described in \cref{subsec:reduction-to-assertion}).
This assertion problem can then be solved by arbitrary off-the-shelf solvers such as bounded model checkers and CHC solvers~\cite{horn1951sentences,bjorner2015horn}, as long as they accept LLVM bitcode input.
(A CHC solver SeaHorn was adopted as our backend solver; the reason for this choice is explained in \cref{subsec:chc-verification,subsec:synergy-chc}.)
However, naively applying these solvers to the refcount bug detection is infeasible due to the large size and the high complexity of the Linux kernel.
To cope with this problem, \Tool\ takes two additional steps: \emph{modeling the Linux kernel} (described in \cref{subsec:modeling}) and \emph{program slicing} (described in \cref{subsec:slicing}).
The modeling phase replaces specific functions and instructions with simpler abstractions that accurately capture the refcount behavior to mitigate the complexity of the Linux kernel.
The slicing phase removes portions irrelevant to refcount verification in order to reduce the verification size to a manageable level.
With these two steps, \Tool\ achieves fully automated, scalable, and accurate refcount bug detection in Linux drivers.

\subsection{Reduction to Assertion Checking Problem} \label{subsec:reduction-to-assertion}
This section describes how we reduce refcount verification to an assertion checking problem by exploiting the structure of the Linux kernel driver interface.
In general, static refcount bug detection in the Linux kernel tends to lack scalability, accuracy, or both due to its vastness and complexity.
However, refcount invariants in drivers are simpler than those in the entire kernel.
The kernel provides data structures such as \code{struct file_operations}, \code{struct net_device_ops}, and \code{struct platform_driver}, and device drivers implement instances of these structures based on the type of devices.
Each data structure contains function pointers that are invoked by the kernel when certain events occur.
In the initialization phase, for example, the kernel calls \code{open()} for \code{struct file_operations}, \code{ndo_init()} for \code{struct net_device_ops}, and \code{probe()} for \code{struct platform_driver}.
To prevent memory leaks, all acquired resources must be released when the initialization fails.
Based on this observation, we derive a refcount invariant applicable to all drivers: \textit{the refcount difference should be zero when initialization fails}.
We verify this invariant by automatic reduction to an assertion checking problem.
Figure~\ref{fig:drvhorn-overview} abstracts the generated code for the assertion problem.
It calls the driver's initialization function and asserts whether all refcount differences are zero upon failure.
The modeling of device, device tree node, and firmware node refcounts is detailed in \cref{subsec:modeling}.

\begin{figure}[ht]
\begin{minted}[escapeinside=||]{c}
int main(void) {
  int err = <initialization function>();
  if (err) {
    assert(|$\Delta$|(refcount of devices) == 0);
    assert(|$\Delta$|(refcount of device tree nodes) == 0);
    assert(|$\Delta$|(refcount of firmware nodes) == 0);
    ...
  }
  return 0;
}
\end{minted}
\caption{Abstract overview of the verification target.}
\label{fig:drvhorn-overview}
\end{figure}

The assertion checking problem is expressed in LLVM bitcode~\cite{lattner2004llvm} (Figure~\ref{fig:drvhorn-overview} is written in C-like syntax solely for readability).
We chose LLVM bitcode as the intermediate representation because it is suited to modeling low-level details of drivers.
This choice also allows us to leverage various existing tools that accept LLVM bitcode as input, such as SeaHorn and SeaBMC~\cite{priya2022bounded}.
However, these solvers do not support LLVM's inline assembly instructions, which are commonly used in drivers.
To address this limitation, we replace inline assembly instructions with equivalent LLVM IR instructions in terms of the refcount invariant.
Currently \Tool\ supports 105 kinds of x86-64 inline assembly instructions, covering all inline assembly instructions encountered in our evaluation.
For unsupported ones, \Tool\ replaces them with non-deterministic values.
This may cause false positives, but we expect minimal impact because our slicing typically removes most inline assembly instructions.

\subsection{Linux Kernel Modeling} \label{subsec:modeling}
\Tool\ models key Linux kernel APIs to preserve refcount behavior while drastically reducing verification complexity; users do not need to implement any modeling themselves.
We address the main categories of modeled APIs below.

\subsubsection{Device Tree} \label{subsubsec:modeling_device_tree}
The device tree~\cite{linux_dt} describes hardware components in a tree structure, where each node is represented by a \code{struct device_node} which is a refcounted object.
Since each node has pointers to other nodes, such as its parent, children, and siblings, direct translation to LLVM bitcode entails complicated pointer analysis, leading to time-consuming and inaccurate verification.
We address this issue by modeling the device tree APIs~\cite{linux_dt_api} rather than the actual tree structure.

Modeling the device tree APIs requires only minimal implementation effort by noting common patterns.
All device tree APIs can be classified into four types, summarized in Table~\ref{tab:dt-api-types}.
As APIs of the same type are modeled in the same way, keeping up with the new API changes will be straightforward since the only required effort is to determine which type the new API belongs to.

\begin{table}[H]
\centering
\caption{Four types of device tree APIs and their examples}
\label{tab:dt-api-types}
\begin{tabular}{|c|c|c|}
\hline
& \textbf{Returns NULL*} & \textbf{Returns non-NULL*} \\
\hline
\textbf{Decrements RC**} & \code{of_get_next_child} & \code{of_find_node_by_name} \\
& \code{of_get_next_available_child} & \code{of_find_node_by_type} \\
\hline
\textbf{Does not} & \code{of_get_parent} & \code{of_find_next_cache_node} \\
\textbf{decrement RC**} & \code{of_get_child_by_name} & \code{of_graph_get_port_by_id} \\
\hline
\end{tabular}

\vspace{2mm}
\small{*when passed NULL as input. **RC = refcount of given node.}
\end{table}

\subsubsection{Devices and Buses} \label{subsubsec:modeling_devices_buses}
Many kernel APIs, such as \code{bus_find_device()}, return a refcount-incremented \code{struct device} embedded in a bus-specific object (for example, \code{struct usb_interface}).
If we pass these APIs directly to the backend solver, the returned device pointer is unconstrained and may alias unrelated objects, leading to spurious refcount violations.
We therefore model such APIs so that each call returns a device embedded in a specific object determined by the bus argument, whose memory region does not alias other objects, as illustrated by our modeling of \code{usb_find_interface()} in Figure~\ref{fig:usb_find_interface}.

\begin{figure}[H]
\begin{minted}[escapeinside=||]{c}
const struct bus_type usb_bus_type = ...;
struct usb_interface *usb_find_interface(...) {
  struct device *dev = bus_find_device(&usb_bus_type, ...);
  return dev ? container_of(dev, struct usb_interface, dev) : NULL;
}
\end{minted}
\cprotect\caption{Abstract implementation of \code{usb_find_interface()}.}
\label{fig:usb_find_interface}
\end{figure}

In this example, \code{bus_find_device()} returns a refcount-incremented \code{struct device} embedded in a \code{struct usb_interface}, and the driver recovers the enclosing interface structure via \code{container_of}.

\subsubsection{Device Resource Management} \label{subsubsec:modeling_devres}
Devres provides a device-managed layer of cleanup.
\code{__devm_add_action()} is the API that registers a devres entry by associating a cleanup function with a device.
In our model we do not track the devres list explicitly; instead, we replace the return value of \code{__devm_add_action()} with \code{0} (success) and immediately insert a call to the given cleanup function after the call site.
This simplification preserves the effect of releasing resources while keeping the encoding small for the backend solver.

\subsection{Program Slicing} \label{subsec:slicing}
Reducing the code size is required to mitigate the Linux kernel's vastness.
A simple approach is to use LLVM's Dead Code Elimination (DCE) passes~\cite{llvm_passes}, which remove unnecessary instructions and functions.
However, DCE passes alone are insufficient to reduce the code size to a manageable level, as shown in \cref{subsec:eval-main}.
Therefore, we implement a more aggressive slicing technique.

The slicing consists of two steps.
The first step is to collect the function arguments that are \emph{essential} for refcount validation.
A function argument is marked as essential if and only if it embeds \code{struct kref} and the function body can modify the refcount.
For example, arguments of \code{get_device()} and \code{put_device()} are marked as essential because they embed \code{struct kref} and the function body can modify the refcount.
The second step involves collecting necessary instructions for each function.
The necessity of an instruction is determined based on the following conditions:
(here, $P$ is considered an \emph{underlying object} of $Q$ when $P$ is obtained by \code{llvm::getUnderlyingObject(Q)} or similar LLVM API calls on $Q$.
This means that when tracing back $Q$ through pointer arithmetic and casts, $P$ is identified as the original memory allocation or variable.)
\begin{itemize}
  \item It is a \code{br}, \code{ret}, \code{phi}, \code{switch}, or \code{alloca}.
  \item It is a \code{call} whose parameter's underlying object is an essential argument.
  \item It is a \code{load} reading from a memory location whose underlying object is the result of a necessary instruction.
  \item It is a \code{store} writing to a memory read by a necessary \code{load} instruction.
  \item All its operands are necessary instructions.
\end{itemize}

This process starts from each function's \code{call} and \code{ret}, recursively collecting necessary instructions.
If an instruction is deemed necessary based on the above conditions, then the same process is applied to the terminator instructions of the predecessor basic blocks.
Instructions marked as necessary are retained, and the rest are removed.
If an operand of a retained instruction is removed, it is replaced with a non-deterministic value of the same type.
Replacing function calls with non-deterministic values may drop critical side effects, and we mitigate this problem by filling non-deterministic values to arguments with LLVM's \code{writeonly} attribute.

An example of how the slicing works is shown in Figure~\ref{fig:slicing-example}.
This is a simplified version of an actual code of \code{tpm_bios_measurements_open()} described in Figure~\ref{fig:tool-overview}.
Instructions that are essential for our refcount verification, such as \code{get_device()} and \code{put_device()} are retained as-is, while non-essential instructions such as \code{inode_lock} and \code{inode_unlock}, highlighted in red, are removed.
Note that \code{seq_open()} is replaced with a call to a non-deterministic function, as the function body is not relevant to refcount verification while its result is used to determine whether to call \code{put_device()}.

\begin{figure}[h]
\centering
\begin{minipage}[t]{0.45\textwidth}
\begin{minted}[escapeinside=@@]{c}
static int
tpm_bios_measurements_open(...) {
  @\colorbox{highlightred}{inode\_lock(inode);}@
  get_device(&chip->dev);
  @\colorbox{highlightred}{inode\_unlock(inode);}@
  err = @\colorbox{highlightred}{seq\_open(...)}@;
  if (err) put_device(&chip->dev);
  return err;
}
\end{minted}
\end{minipage}
\hfill
\begin{minipage}[t]{0.45\textwidth}
\begin{minted}{c}
extern int nondet(void);
static int
tpm_bios_measurements_open(...) {
  get_device(&chip->dev);
  err = nondet();
  if (err) put_device(&chip->dev);
  return err;
}
\end{minted}
\end{minipage}
\cprotect\caption{Simplified version of \code{tpm_bios_measurements_open()} before (left) and after (right) slicing.}
\label{fig:slicing-example}
\end{figure}

\subsection{CHC-based Verification} \label{subsec:chc-verification}
As described in \cref{subsec:reduction-to-assertion}, it suffices to solve the assertion checking problem to verify the refcount invariant.
Arbitrary off-the-shelf solvers can be used to solve this problem if they can accept LLVM bitcode as the input.
We adopted SeaHorn~\cite{gurfinkel2015seahorn} as our backend solver since it is suited for verifying programs with loops, which are common in refcount manipulations in the Linux kernel as can be seen in Figure~\ref{fig:example-refcount-bug}.
After slicing, \Tool\ reduces much of the remaining interprocedural structure by using SeaHorn's inlining option.
Calls that remain, such as recursive calls, are handled by SeaHorn's CHC-based verification.
We do not add extra preprocessing specifically for interprocedural analysis beyond the slicing described above.
We address the effectiveness of using CHC-based verification by comparing it with a bounded model checker in \cref{subsec:synergy-chc}.

\section{Evaluation} \label{sec:evaluation}
We evaluated \Tool\ by verifying all \totalPlatformDrivers\ platform drivers in x86-64 Linux kernel v6.6, compiled with \code{allyesconfig}.
Although \Tool\ can target other types of drivers, such as \code{file_operations} and \code{i2c_driver}, we chose platform drivers as the primary evaluation target due to their large number (\totalPlatformDrivers\ vs. 1017 \code{file_operations} and 1067 \code{i2c_driver}), which provides better statistical significance for the evaluation.
Our evaluation was done on an Ubuntu 24.04 machine with sixteen 2.30 GHz Intel Core i9 x86-64 cores and 64 GB of RAM.
We set a 300-second timeout for each driver verification.

\subsection{Overall Results} \label{subsec:eval-main}
Table~\ref{tab:overall-results} summarizes the overall results of our evaluation.
Out of \totalPlatformDrivers\ platform drivers, \Tool\ timed out on \timeouts\ and reported refcount bugs in \allBugs\ of them.
Manual inspection confirmed \realBugs\ as real bugs, resulting in \falsePositiveCount\ (\falsePositiveRate) false positives.
Among the confirmed bugs, \unknownBugs\ were previously unknown, and \mergedPatches\ of our submitted patches were merged into the Linux kernel.
Out of the \falsePositiveCount\ false positives, \falsePositiveBySlicing\ were caused by the unsoundness of our slicing policy.

\begin{table}[ht]
\centering
\caption{Overall evaluation results on x86-64 Linux v6.6 platform drivers}
\label{tab:overall-results}
\begin{tabular*}{\columnwidth}{@{\extracolsep{\fill}}lr}
\hline
\textbf{Metric} & \textbf{Value} \\
\hline
Total drivers analyzed & \totalPlatformDrivers \\
\hline
Bugs reported & \allBugs \\
Real bugs & \realBugs \\
False positives & \falsePositiveCount (\falsePositiveRate) \\
Previously unknown bugs & \unknownBugs \\
\hline
Patches merged & \mergedPatches \\
\hline
Drivers timed out (300s) & \timeouts \\
\hline
\end{tabular*}
\end{table}

We also tested the effectiveness of our slicing policy by comparing it with LLVM's DCE passes~\cite{llvm_passes}.
For the \realBugs\ drivers in which \Tool\ found real refcount bugs, we ran the verification with DCE passes enabled but with our program slicing disabled.
As a result, all the drivers exceeded the timeout limit of the verification.
This is because sound slicing techniques such as DCE passes tend to be too conservative, leaving most of the code intact and leading to a state space explosion.
\subsection{Kernel Bug Fixes and Contributions} \label{subsec:categories}
We classify the detected bugs into three categories: memory leak fixes, UAF fixes, and API misuse fixes.
We briefly illustrate each category; simplified code examples are
\iffull
deferred to the appendix.
\else
deferred to the longer version~\cite{CAV26-full}.
\fi
The most common category is memory leaks: \memLeakPatches\ of our \mergedPatches\ accepted patches fix missing decrements on error paths, such as the added \code{put_device()} in \code{tpm_bios_measurements_open()} and the newly introduced cleanup path
\iffull
in \code{q6v5_wcss_probe()} (Appendix~\ref{appendix:tpm-leak} and Appendix~\ref{appendix:q6v5}).
else
in \code{q6v5_wcss_probe()}.
\fi
We found a UAF bug caused by a misunderstanding of device tree APIs: the driver calls \code{of_find_node_by_name()} and then \code{of_node_put()} on \code{dev->of_node} in a loop even though the former already decrements the node's refcount; the accepted fix for this specific case rewrites the loop to use \code{for_each_child_of_node()} and \code{of_get_child_by_name()} and removes the extra
\iffull
\code{of_node_put()} (Appendix~\ref{appendix:tegra-uaf}).
\else
\code{of_node_put()}.
\fi
API misuse fixes cover cases where our modeling of some kernel APIs uncovers incorrect usage patterns: in commit 54a8cd0, the model exposes that the pointer returned by \code{device_link_add()} with \code{DL_FLAG_AUTOREMOVE_SUPPLIER} should only be used for null checks, so we remove the stored \code{struct device_link *link} field from \code{struct qcom_ice} and treat the return value as a boolean indicator
\iffull
instead (Appendix~\ref{appendix:devlink-misuse}).
\else
instead.
\fi
In addition to these bug fixes, we also contributed a new general API \code{devm_device_init_wakeup()}.
Wakeup devices allow the system to resume from low-power states, and the existing \code{device_init_wakeup()} API initializes a wakeup device but requires developers to manually pair each call with \code{device_disable_wakeup()} in all error and cleanup paths.
\Tool\ revealed that many drivers lacked this cleanup, making all of them leaking wakeup devices.
Instead of fixing each driver separately, we introduced \code{devm_device_init_wakeup()}, a device-managed variant that automatically disables wakeup when the device is detached, so developers can simply replace \code{device_init_wakeup()} with the new API.
This change was merged as commit b3172683 and is now used by \devmWakeupUsers\ drivers.

\subsection{Evaluation of a CHC Solver as the Backend Verifier} \label{subsec:synergy-chc}

To evaluate whether CHC-based verification is well-suited for detecting refcount bugs, we conducted a comparative experiment with SeaBMC~\cite{priya2022bounded}, the bounded model checking option of SeaHorn.
For the \realBugs\ drivers in which \Tool\ found real refcount bugs, SeaBMC missed \bmcErrorCountInDetected\ (\bmcErrorRateInDetected) and timed out on \bmcTimeoutCountInDetected\ (\bmcTimeoutRateInDetected), including the bug in Figure~\ref{fig:example-refcount-bug}.
We ran SeaBMC with loop-unrolling bound 1, following prior Linux kernel model-checking work, because larger bounds timed out for most drivers in our setting.
Although we did not perform a precise per-bug classification, insufficient loop unrolling explains many of SeaBMC's missed bugs, including the bug in Figure~\ref{fig:example-refcount-bug}.
For the \timeouts\ drivers where \Tool\ timed out, SeaBMC also timed out on \bmcTimeoutInTimeout\ and reported \bmcBugsInTimeout\ refcount bugs, of which only \bmcCorrectBugsInTimeout\ were real and their root causes had already been addressed by our patches.
Overall, these results indicate that CHC-based verification is more effective than bounded model checking for our purpose of automatically finding refcount bugs in Linux drivers.

\subsection{Analysis on False Positives, False Negatives, and Timeouts} \label{subsec:false-positives}

Our slicing policy does not consider indirect information flow.
However, only \falsePositiveBySlicing\ out of the \falsePositiveCount\ false positives were caused by this unsoundness, demonstrating the effectiveness of our slicing policy.
One example of false positive is the
\iffull
\code{acer_platform_probe()} described in Appendix~\ref{appendix:acer-fp},
\else
\code{acer_platform_probe()} given in \cite{CAV26-full},
\fi
where the real driver correctly balances initialization and cleanup, but the sliced program may not.
The remaining \falsePositiveByMemoryAnalysis\ were caused by conservative pointer analysis inherited from SeaHorn's memory model~\cite{gurfinkel2015seahorn,gurfinkel2017context}, where certain aliasing patterns are over-approximated to keep verification scalable.

To approximate the false negative rate, we tested whether \Tool\ could detect known refcount bugs since the initial Linux v6.6 release.
We scanned commit messages in the \code{drivers} directory for refcount-related keywords\footnote{We used ``refcount leak'', ``reference leak'', ``refcount bug'', ``reference count bug'', ``reference count leak'', ``refcount imbalance'', ``reference count imbalance'', ``node leak'', ``leaked reference'', ``leaked node'', ``leaked OF node'', ``leaks reference'', ``leaks node'', and ``leaks OF node''}, and found 22 bug-fix commits, 17 of which fixed bugs present in the x86-64 \code{allyesconfig} kernel.
Out of these 17 bugs, \Tool\ successfully detected 14, timed out on two, and missed one due to the unsoundness of our slicing policy, suggesting a low false negative rate despite the small sample size.
The missed bug was fixed by commit d029ede: our slicing did not account for indirect information flow through \code{fwnode_property_read_u32()}, so the store to \code{port_num} was treated as unnecessary and the erroneous call to \code{qca8k_parse_port_leds()} became invisible to the verifier.

Drivers with multiple/nested loops tend to time out.
For example, \timeoutByMtkClk\ timeouts were caused by a kernel utility function \code{__mtk_clk_simple_probe()}.
This function calls a number of \code{mtk_clk_register_*} functions that update multiple refcounts in loops, and most of its instructions survived slicing.
The resulting CHC problem requires complex loop/refcount invariants, which can cause the solver to diverge or time out.
Abstract interpretation could be viewed as a complementary technique for improving CHC solving or for supporting slicing, rather than as a \Tool-specific replacement.

\section{Related Work} \label{sec:related-work}
Previous work has explored static bug detection in OS drivers.
However, fully automated and scalable Linux driver verification with high accuracy has been challenging due to its diversity, as opposed to the more uniform Windows driver framework.
\textsc{Slam}~\cite{ball2011decade} verifies conformance to API usage rules of Windows drivers, which formed the basis of Microsoft's SDV~\cite{ball2006thorough}, a static analyzer for Windows drivers.
Avinux~\cite{post2007integrated,post2009towards} adapts SDV for verifying Linux drivers.
It identified several known and six previously unknown bugs in Linux drivers, but considerable amount of manual effort was needed to address Linux's flexible and diverse driver framework, making it not suitable for our goal of fully automated and scalable refcount bug detection in Linux drivers.
\textsc{Blast}~\cite{henzinger2003software} discovered several bugs in Linux drivers and Windows drivers.
However, it would face challenges in large scale automated verification for Linux drivers, as pointed out by Mühlberg et al.~\cite{muhlberg2006blasting}:

\begin{quote}
  \emph{``In our experience, Blast is rather difficult to apply by a practitioner during OS software development. [...]
    Especially in the case of memory safety properties, massive changes to the source code were necessary which essentially requires one to know about a bug beforehand.''}
\end{quote}

While static refcount bug detection in the Linux kernel has been previously studied, the complexity of the Linux kernel makes deriving precise refcount invariants challenging, as different subsystems follow different resource management patterns.
RID~\cite{mao2016rid} detects bugs by comparing execution paths in the same function with identical return values, reporting discrepancies in refcount changes.
While RID identified 83 bugs in the DPM subsystem, it resulted in a false positive rate of 76.6\%, and does not generalize well to other subsystems as it focuses on verifying specific API usages in DPM.
CID~\cite{tan2021detecting} and LinKRID~\cite{liu2022linkrid} extend the scope of refcount analysis using symbolic execution, with LinKRID detecting 118 bugs across the kernel.
However, the challenge of introducing accurate invariants across diverse kernel subsystems leads to false positive rates of 70.5\% and 43.5\%, respectively, which may limit practical adoption.

In contrast, \Tool\ addresses this challenge by focusing on drivers and exploiting the kernel-provided driver interface to derive a simple, universal invariant.
This enables effective kernel modeling and aggressive program slicing, achieving both high accuracy and scalability.

\section{Conclusion and Future Work} \label{sec:conclusion}
We presented \Tool, a static analysis tool for automated refcount bug detection in Linux drivers.
We introduced a refcount invariant: the refcount difference should be zero when the driver initialization fails.
This invariant is applicable to all drivers and enables scalable and accurate detection.
Our evaluation on all \totalPlatformDrivers\ platform drivers of the x86-64 Linux v6.6 demonstrates the scalability and accuracy of \Tool.
\Tool\ reported \allBugs\ bugs, out of which \realBugs\ were real, showing a relatively low false positive rate of \falsePositiveRate.
\unknownBugs\ of them were previously unknown, and \mergedPatches\ patches by us were merged into the Linux kernel to address the root causes.
We also evaluated false negatives by testing whether \Tool\ was able to detect known refcount bugs since the initial v6.6 release and found that \Tool\ missed only one bug out of 17.
Verifying refcounts other than \code{struct kref} and
extending the invariant to cover bugs during and after successful driver initialization are left for future work.

\subsubsection*{\ackname}
This work was supported by JSPS KAKENHI Grant Number JP20H05703 and JP26H02486.

\subsubsection*{Disclosure of Interests}
All authors declare that they have no competing interests.

\subsubsection*{Data-Availability Statement}
All experimental results can be reproduced by the package \url{https://zenodo.org/records/20082488}

\bibliographystyle{splncs04}
\bibliography{ref}

@inproceedings{gurfinkel2015seahorn,
  title={The {SeaHorn} verification framework},
  author={Gurfinkel, Arie and Kahsai, Temesghen and Komuravelli, Anvesh and Navas, Jorge A},
  booktitle={International Conference on Computer Aided Verification},
  pages={343--361},
  year={2015},
  organization={Springer}
}

@inproceedings{priya2022bounded,
  title={Bounded model checking for {LLVM}},
  author={Priya, Siddharth and Su, Yusen and Bao, Yuyan and Zhou, Xiang and Vizel, Yakir and Gurfinkel, Arie},
  booktitle={2022 Formal Methods in Computer-Aided Design (FMCAD)},
  pages={214--224},
  year={2022},
  organization={TU Wien Academic Press}
}

@inproceedings{mao2016rid,
  title={{RID}: finding reference count bugs with inconsistent path pair checking},
  author={Mao, Junjie and Chen, Yu and Xiao, Qixue and Shi, Yuanchun},
  booktitle={Proceedings of the Twenty-First International Conference on Architectural Support for Programming Languages and Operating Systems},
  pages={531--544},
  year={2016}
}

@inproceedings{tan2021detecting,
  title={Detecting kernel refcount bugs with $\{$Two-Dimensional$\}$ consistency checking},
  author={Tan, Xin and Zhang, Yuan and Yang, Xiyu and Lu, Kangjie and Yang, Min},
  booktitle={30th USENIX Security Symposium (USENIX Security 21)},
  pages={2471--2488},
  year={2021}
}

@inproceedings{liu2022linkrid,
  title={{LinKRID}: Vetting Imbalance Reference Counting in Linux kernel with Symbolic Execution},
  author={Liu, Jian and Yi, Lin and Chen, Weiteng and Song, Chengyu and Qian, Zhiyun and Yi, Qiuping},
  booktitle={31st USENIX Security Symposium (USENIX Security 22)},
  pages={125--142},
  year={2022}
}

@inproceedings{post2007integrated,
  title={Integrated static analysis for Linux device driver verification},
  author={Post, Hendrik and K{\"u}chlin, Wolfgang},
  booktitle={International conference on integrated formal methods},
  pages={518--537},
  year={2007},
  organization={Springer}
}

@inproceedings{emmi2009verifying,
  title={Verifying reference counting implementations},
  author={Emmi, Michael and Jhala, Ranjit and Kohler, Eddie and Majumdar, Rupak},
  booktitle={International Conference on Tools and Algorithms for the Construction and Analysis of Systems},
  pages={352--367},
  year={2009},
  organization={Springer}
}

@incollection{bjorner2015horn,
  title={Horn clause solvers for program verification},
  author={Bj{\o}rner, Nikolaj and Gurfinkel, Arie and McMillan, Ken and Rybalchenko, Andrey},
  booktitle={Fields of Logic and Computation II: Essays Dedicated to Yuri Gurevich on the Occasion of His 75th Birthday},
  pages={24--51},
  year={2015},
  publisher={Springer}
}

@inproceedings{muhlberg2006blasting,
  title={Blasting linux code},
  author={M{\"u}hlberg, Jan Tobias and L{\"u}ttgen, Gerald},
  booktitle={International Workshop on Parallel and Distributed Methods in Verification},
  pages={211--226},
  year={2006},
  organization={Springer}
}

@inproceedings{henzinger2003software,
  title={Software verification with {BLAST}},
  author={Henzinger, Thomas A and Jhala, Ranjit and Majumdar, Rupak and Sutre, Gregoire},
  booktitle={Model Checking Software: 10th International SPIN Workshop Portland, OR, USA, May 9--10, 2003 Proceedings 10},
  pages={235--239},
  year={2003},
  organization={Springer}
}

@inproceedings{lattner2004llvm,
  title={{LLVM}: A compilation framework for lifelong program analysis \& transformation},
  author={Lattner, Chris and Adve, Vikram},
  booktitle={International symposium on code generation and optimization, 2004. CGO 2004.},
  pages={75--86},
  year={2004},
  organization={IEEE}
}

@inproceedings{gurfinkel2017context,
  title={A context-sensitive memory model for verification of {C}/{C++} programs},
  author={Gurfinkel, Arie and Navas, Jorge A},
  booktitle={Static Analysis: 24th International Symposium, SAS 2017, New York, NY, USA, August 30--September 1, 2017, Proceedings 24},
  pages={148--168},
  year={2017},
  organization={Springer}
}

@inproceedings{kroah2004kobjects,
  title={kobjects and krefs},
  author={Kroah-Hartman, Greg},
  booktitle={Linux Symposium},
  pages={295},
  year={2004}
}

@inproceedings{bai2024countdown,
  title={CountDown: Refcount-guided Fuzzing for Exposing Temporal Memory Errors in Linux Kernel},
  author={Bai, Shuangpeng and Zhang, Zhechang and Hu, Hong},
  booktitle={Proceedings of the 2024 on ACM SIGSAC Conference on Computer and Communications Security},
  pages={1315--1329},
  year={2024}
}

@article{horn1951sentences,
  title={On sentences which are true of direct unions of algebras1},
  author={Horn, Alfred},
  journal={The Journal of Symbolic Logic},
  volume={16},
  number={1},
  pages={14--21},
  year={1951},
  publisher={Cambridge University Press}
}

@article{ball2006thorough,
  title={Thorough static analysis of device drivers},
  author={Ball, Thomas and Bounimova, Ella and Cook, Byron and Levin, Vladimir and Lichtenberg, Jakob and McGarvey, Con and Ondrusek, Bohus and Rajamani, Sriram K and Ustuner, Abdullah},
  journal={ACM SIGOPS Operating Systems Review},
  volume={40},
  number={4},
  pages={73--85},
  year={2006},
  publisher={ACM New York, NY, USA}
}

@article{post2009towards,
  title={Towards automatic software model checking of thousands of Linux modules—A case study with {Avinux}},
  author={Post, Hendrik and Sinz, Carsten and K{\"u}chlin, Wolfgang},
  journal={Software Testing, Verification and Reliability},
  volume={19},
  number={2},
  pages={155--172},
  year={2009},
  publisher={Wiley Online Library}
}

@article{ball2011decade,
  title={A decade of software model checking with {SLAM}},
  author={Ball, Thomas and Levin, Vladimir and Rajamani, Sriram K},
  journal={Communications of the ACM},
  volume={54},
  number={7},
  pages={68--76},
  year={2011},
  publisher={ACM New York, NY, USA}
}

@misc{cve-2019-11487,
  title={{MITRE. CVE-2019-11487}},
  note={\url{https://cve.mitre.org/cgi-bin/cvename.cgi?name=CVE-2019-11487}},
}

@misc{cve-2023-7192,
  title={{MITRE. CVE-2023-7192}},
  note={\url{https://cve.mitre.org/cgi-bin/cvename.cgi?name=CVE-2023-7192}},
}

@misc{llvm_passes,
  title={{{LLVM}'s Analysis and Transform Passes}},
  note={\url{https://llvm.org/docs/Passes.html}},
}

@misc{linux_dt,
  title={{Linux and the Devicetree}},
  note={\url{https://docs.kernel.org/devicetree/usage-model.html}},
}

@misc{linux_dt_api,
  title={{Devicetree Kernel API}},
  note={\url{https://docs.kernel.org/devicetree/kernel-api.html}}
}

@misc{CAV26-full,
      title={Automatic Detection of Reference Counting Bugs in Linux Kernel Drivers}, 
      author={Joe Hattori and Naoki Kobayashi and Ken Sakayori},
      year={2026},
      eprint={2605.13246},
      archivePrefix={arXiv},
      primaryClass={cs.CR},
      url={https://arxiv.org/abs/2605.13246}, 
}

\iffull
\clearpage
\appendix
\section*{Appendix}
\section{Patch}
\subsection{Commit 5d8e297} \label{appendix:tpm-leak}
Related to \cref{subsec:categories} (Kernel Bug Fixes and Contributions).
This example fixes a missing error-path \code{put_device()} in \code{tpm_bios_measurements_open()} to balance refcounts.

\begin{figure}[h!]
\begin{minted}[escapeinside=||]{c}
int tpm_bios_measurements_open(struct inode *inode, struct file *file) {
  struct tpm_chip *chip = inode->i_private->chip;
  get_device(&chip->dev);
  int err = seq_open(...);
  if (!err) file->private_data->private = chip;
|\diffadd{+ else put\_device(\&chip->dev);}|
  return err;
}
\end{minted}
\caption{Commit 5d8e297}
\label{fig:tpm_bios_measurements_open}
\end{figure}

\code{tpm_bios_measurements_open()} is registered as the \code{open} operation for \code{tpm_bios_measurements_ops}.
It calls \code{get_device()} to increment the refcount of \code{chip->dev}, but the original code did not decrement it in the error path, leading to a memory leak.
The patch adds \code{put_device()} in the error branch to balance the refcount.

\subsection{Commit 60e7c43e} \label{appendix:q6v5}
Related to \cref{subsec:categories} (Kernel Bug Fixes and Contributions).
This example adds error-path cleanup for sub-devices allocated in \code{q6v5_wcss_probe()}.

\begin{figure}[h!]
\begin{minted}[escapeinside=@@]{c}
static int q6v5_wcss_probe(struct platform_device *pdev) {
  struct q6v5_wcss *wcss;
  ...
  ret = qcom_q6v5_init(...);
  if (ret)
    return ret;

  qcom_add_glink_subdev(rproc, &wcss->glink_subdev, "q6wcss");
  qcom_add_pdm_subdev(rproc, &wcss->pdm_subdev);
  qcom_add_ssr_subdev(rproc, &wcss->ssr_subdev, "q6wcss");

  if (desc->ssctl_id) {
    wcss->sysmon = qcom_add_sysmon_subdev(...);
@\diffadd{+   if (IS\_ERR(wcss->sysmon)) \{}@
@\diffadd{+     ret = PTR\_ERR(wcss->sysmon);}@
@\diffadd{+     goto deinit\_remove\_subdevs;}@
@\diffadd{+   \}}@
  }
  ret = rproc_add(rproc);
  if (ret)
@\diffrem{-    return ret;}@
@\diffadd{+    goto remove\_sysmon\_subdev;}@
  ...
  return 0;
@\diffadd{+remove\_sysmon\_subdev:}@
@\diffadd{+ if (desc->ssctl\_id)}@
@\diffadd{+   qcom\_remove\_sysmon\_subdev(wcss->sysmon);}@

@\diffadd{+deinit\_remove\_subdevs:}@
@\diffadd{+ qcom\_q6v5\_deinit(\&wcss->q6v5);}@
@\diffadd{+ qcom\_remove\_glink\_subdev(rproc, \&wcss->glink\_subdev);}@
@\diffadd{+ qcom\_remove\_pdm\_subdev(rproc, \&wcss->pdm\_subdev);}@
@\diffadd{+ qcom\_remove\_ssr\_subdev(rproc, \&wcss->ssr\_subdev);}@
@\diffadd{+ return ret;}@
}
\end{minted}
\caption{Commit 60e7c43e}
\label{fig:q6v5_wcss}
\end{figure}

\code{q6v5_wcss_probe()}, the initialization function of the \code{q6v5_wcss} driver, adds several sub-devices by calling functions like \code{qcom_add_glink_subdev()}, \code{qcom_add_pdm_subdev()}, \code{qcom_add_ssr_subdev()}, and \code{qcom_add_sysmon_subdev()}.
However, these allocated sub-devices were not released in the error path, leading to memory leaks.
Figure~\ref{fig:q6v5_wcss} shows a simplified version of our accepted patch 60e7c43e.
It adds the full error handling code to fix this bug.

\subsection{Commit b9784e5cd} \label{appendix:tegra-uaf}
Related to \cref{subsec:categories} (Kernel Bug Fixes and Contributions).
This example fixes a device tree refcounting misuse that could lead to a use-after-free.

\begin{figure}[h!]
\begin{minted}[escapeinside=@@]{c}
#define EMC_LABEL "emc-tables"
#define LPDDR2 "lpddr2"
static struct device_node *tegra_emc_find_node_by_ram_code(
    struct tegra_emc *emc) {
  struct device *dev = emc->dev;
  struct device_node *np;
  ...
@\diffrem{- for (np = of\_find\_node\_by\_name(dev->of\_node, EMC\_LABEL); np; }@
@\diffrem{-      np = of\_find\_node\_by\_name(np, EMC\_LABEL)) \{}@
@\diffadd{+ for\_each\_child\_of\_node(dev->of\_node, np) \{}@
@\diffadd{+   if (!of\_node\_name\_eq(np, EMC\_LABEL)) continue;}@
    ...
    bool cfg_mismatches;
@\diffrem{-   struct device\_node *lpddr2\_np = of\_find\_node\_by\_name(np, LPDDR2);}@
@\diffadd{+   struct device\_node *lpddr2\_np = of\_get\_child\_by\_name(np, LPDDR2);}@
    ...
    if (cfg_mismatches)
@\diffrem{-     of\_node\_put(np);}@
      continue;
    ...
  }
  ...
}
\end{minted}

\caption{Commit b9784e5cd}
\label{fig:tegra_uaf}
\end{figure}

The original code passes \code{dev->of_node} to \code{of_find_node_by_name()} and then calls \code{of_node_put()} inside the loop, not accounting for the fact that \code{of_find_node_by_name()} already decrements the refcount of the given node.
This can lead to use-after-free bugs.
The patch replaces the loop with \code{for_each_child_of_node()} and \code{of_get_child_by_name()}, and removes the extra \code{of_node_put()}.

\subsection{Commit 54a8cd0} \label{appendix:devlink-misuse}
Related to \cref{subsec:categories} (Kernel Bug Fixes and Contributions).
This example removes a stored device link pointer that should only be used for a null check.

\begin{figure}[h!]
\begin{minted}[escapeinside=@@]{c}
struct qcom_ice {
  ...
@\diffrem{- struct device\_link *link;}@
};

struct qcom_ice *of_qcom_ice_get(struct device *dev) {
  struct qcom_ice *ice;
@\diffadd{+ struct device\_link *link;}@
  ...
@\diffrem{- ice->link = device\_link\_add(dev, ..., DL\_FLAG\_AUTOREMOVE\_SUPPLIER);}@
@\diffrem{- if (!ice->link)}@
@\diffadd{+ link = device\_link\_add(dev, ..., DL\_FLAG\_AUTOREMOVE\_SUPPLIER);}@
@\diffadd{+ if (!link)}@
  ...
}
\end{minted}

\caption{Commit 54a8cd0}
\label{fig:devlink}
\end{figure}

With the \code{DL_FLAG_AUTOREMOVE_SUPPLIER} flag, the device link created by \code{device_link_add()} is automatically managed by the driver core, and the returned pointer is only intended for null checks.
The original code stored this pointer in a field \code{struct device_link *link;} of \code{struct qcom_ice}, which was unnecessary.
The patch removes the field and uses a local variable solely for the null check.

\section{Additional Examples}
\subsection{False Positive Example: \code{acer_platform_probe}} \label{appendix:acer-fp}
Related to \cref{subsec:false-positives} (Analysis on False Positives, False Negatives, and Timeouts).
This example has balanced cleanup paths, but our sliced program can obscure that balance and yield a false positive.

\begin{figure}[h!]
\begin{minted}[escapeinside=@@]{c}
static int acer_platform_probe(struct platform_device *device) {
  int err;
  if (has_cap(ACER_CAP_MAILLED)) {
    err = acer_led_init(&device->dev);
    if (err) goto error_mailled;
  }
  ...
  if (err) goto error_brightness;
  return err;
error_brightness:
  if (has_cap(ACER_CAP_MAILLED)) acer_led_exit();
error_mailled:
  return err;
}

\end{minted}

\cprotect\caption{Implementation of \code{acer_platform_probe}.}
\label{fig:acer}
\end{figure}

\section{Paper-Code Mapping}
This section briefly explains where to look in the codebase.
As already mentioned, \Tool{} is availabe at \url{github.com/joehattori/drvhorn}.
\Tool{} is developed as a fork of SeaHorn, and most of the modifications are made in the directory~\href{https://github.com/joehattori/drvhorn/tree/main/lib/Transforms/Kernel}{lib/Transforms/Kernel/}.

The following table gives a mapping describing in which file the contents of the paper can be found.
\begin{table}[ht]
\centering
\caption{Correspondence between paper content and code implementation.}
\label{tab:paper-code-mapping}
\begin{tabular}{|c|c|c|}
  \hline
  \textbf{Functionality/Step} & \textbf{Appearance in the paper} &  \textbf{Where it is implemented} \\
  \hline
  Insertion of assertions on the refcount invariants & Section~\ref{subsec:reduction-to-assertion} & \href{https://github.com/joehattori/drvhorn/blob/main/lib/Transforms/Kernel/SetupEntrypoint.cc}{SetupEntrypoint.cc}, \href{https://github.com/joehattori/drvhorn/blob/main/lib/Transforms/Kernel/AssertKrefs.cc}{AssertKrefs.cc} \\
  Modelling of device tree and firmware nodes& Section~\ref{subsec:modeling} & \href{https://github.com/joehattori/drvhorn/blob/main/lib/Transforms/Kernel/Device.cc}{Device.cc} \\
  Modeling of devices, buses, and device classes & Section~\ref{subsec:modeling} & \href{https://github.com/joehattori/drvhorn/blob/main/lib/Transforms/Kernel/Device.cc}{Device.cc} \\
  Modeling of device resource management & Section~\ref{subsec:modeling}  & \href{https://github.com/joehattori/drvhorn/blob/main/lib/Transforms/Kernel/Devm.cc}{Devm.cc} \\
  Modeling of other APIs & Section~\ref{subsec:modeling} &  \href{https://github.com/joehattori/drvhorn/blob/main/lib/Transforms/Kernel/Device.cc}{Device.cc} \\
Program slicing & Section~\ref{subsec:slicing} &  \href{https://github.com/joehattori/drvhorn/blob/main/lib/Transforms/Kernel/Slicer.cc}{Slicer.cc}
 \\
Replacement of inline assemblies & Section~\ref{subsec:reduction-to-assertion} & \href{https://github.com/joehattori/drvhorn/blob/main/lib/Transforms/Kernel/HandleInlineAsm.cc}{HandleInlineAsm.cc} \\

\hline
\end{tabular}
\end{table}

\fi
\end{document}